# Evolution of the electronic excitation spectrum with strongly diminishing hole-density in superconducting Bi$_2$Sr$_2$CaCu$_2$O$_{8+\delta}$

J.W. Alldredge[1], Jinho Lee[1,2], K. McElroy[3], M. Wang[1], K. Fujita[1], Y. Kohsaka[1,4], C. Taylor[1], H. Eisaki[5], S. Uchida[6], P.J. Hirschfeld[7] and J.C. Davis[1,8]

[1] *LASSP, Department of Physics, Cornell University, Ithaca NY 14850 USA*

[2] *School of Physics and Astronomy, University of St. Andrews, North Haugh, St. Andrews, Fife KY16 9SS, Scotland*

[3] *Department of Physics, University of Colorado, Boulder CO, 8030 USA*

[4] *Magnetic Materials Laboratory, RIKEN, Wako 351-0198, JAPAN.*

[5] *NI-AIST, 1-1-1 Central 2, Umezono, Tsukuba, Ibaraki, 305-8568 Japan*

[6] *Department of Physics, University of Tokyo, Tokyo, 113-8656 Japan*

[7] *Department of Physics, University of Florida, Gainesville, FL 32611, USA*

[8] *CMP&MS Department, Brookhaven National Laboratory, Upton, NY 11973, USA*

**A complete knowledge of its excitation spectrum could greatly benefit efforts to understand the unusual form of superconductivity occurring in the lightly hole-doped copper-oxides. Here we use tunnelling spectroscopy to measure the T→0 spectrum of electronic excitations N(E) over a wide range of hole-density *p* in superconducting Bi$_2$Sr$_2$CaCu$_2$O$_{8+\delta}$. We introduce a parameterization for N(E) based upon an anisotropic energy-gap $\Delta(\vec{k}) = \Delta_1 (Cos(k_x) - Cos(k_y))/2$ plus an effective scattering rate which varies linearly with energy $\Gamma_2(E) = \alpha E$. We demonstrate that this form of N(E) allows successful fitting of differential tunnelling conductance spectra throughout much of the Bi$_2$Sr$_2$CaCu$_2$O$_{8+\delta}$ phase diagram. The resulting average Δ$_1$ values rise with falling *p* along the familiar trajectory of excitations to the 'pseudogap' energy, while the key scattering rate $\Gamma_2^* = \Gamma_2(E = \Delta_1)$ increases from below ~1meV to a value approaching 25meV as the system is underdoped from**

**$p\sim16\%$ to $p<10\%$. Thus, a single, particle-hole symmetric, anisotropic energy-gap, in combination with a strongly energy and doping dependent effective scattering rate, can describe the spectra without recourse to another ordered state. Nevertheless we also observe two distinct and diverging energy scales in the system: the energy-gap maximum $\Delta_1$ and a lower energy scale $\Delta_0$ separating the spatially homogeneous and heterogeneous electronic structures.**

Hole-doped copper-oxides have their highest superconducting critical temperature $T_c$ at hole-densities per $CuO_2$ of $p\sim16\%$, and the superconductive state exhibits d-wave symmetry. By measuring STM tip-sample differential conductance $dI/dV(r,V) \equiv g(r,V)$ at each location $r$ and bias voltage V one can achieve energy resolved images of the local-density-of-excitations $N(E)$ because $g(\mathbf{r},V) \propto N(\mathbf{r},E=eV)$ (when the $N(E)$ integrated to the junction formation bias is homogeneous[1]). Near optimal doping, the $g(V)$ spectra appear highly consistent with the theoretical $N(E)$ of a d-wave superconductor; when superconductivity is suppressed by unitary scattering at a Zn atom[2,3] or at the center of a vortex core[3,4], the two particle-hole symmetric peaks in $g(V)$ are also suppressed as expected of the superconducting coherence peaks. Thus there can be little doubt that the measured $N(E)$ near optimal doping is that of the d-wave superconducting state. But as $p$ is reduced, the electronic excitations begin to exhibit[5,6,7] a 'pseudo' gap (PG). This is a momentum-space anisotropic energy gap[5-9] in the excitation spectrum whose effects can be detected by numerous spectroscopic and thermodynamic techniques[6,7] far above the superconducting $T_c$ (which diminishes to zero as $p\rightarrow 0$). The PG energy scale increases linearly with diminishing $p$.

Possible explanations for the PG include, for example, effects of hole-doping an antiferromagnetic Mott insulator[10-14]. Different models for this situation yield an anisotropic energy-gap whose maximum diminishes linearly with increasing $p$ (heuristically, one can view this as a dilution of the antiferromagnetic exchange energy by the holes). But an alternative type of proposal has been that the PG is due to some distinct electronic phase[15,16,17,18] whose anisotropic energy gap represents the breaking of a different symmetry. Measurements solely of the PG energy scale versus $p$ have not resulted in discrimination between these two types of proposals and no consensus exists for the cause of the PG in the electronic excitations of copper-oxides[5,6,7].

A fully detailed knowledge of the T→0 intrinsic spectrum of electronic excitations as a function of doping could help break this impasse. The lifetimes of 'nodal' excitations – those with $k \| (\pi,\pi)$ - have actually been widely studied[19-22]; these states are not the focus of study here. Instead we focus primarily on higher energy excited states which reach all the way to the antinodes $k \sim (\pi,0):(0,\pi)$. Scattering rates for these states have been studied in the superconducting[23] and non-superconducting[24] state at or above optimal doping, revealing strong momentum-space anisotropy of the scattering rate at the Fermi surface. And, using optical techniques Gedik *et al* stimulated these non-nodal excited states and discovered that a dramatic change in their recombination rate occurs near optimal doping[25]. Despite these recent advances, knowledge of the T→0 spectrum of electronic excitations sufficient to constrain the models, does not yet exist.

Here we introduce a new technique for understanding the spatial and doping dependence of the electronic excitation spectrum N(E) of superconducting cuprates. We use single crystals of $Bi_2Sr_2CaCu_2O_{8+\delta}$ (Bi-2212) grown by the floating zone method. Atomically clean and flat surfaces of BiO are achieved and maintained by cleaving the samples in cryogenic ultrahigh vacuum before insertion into the STM at T=4.2K. We report on samples with six different hole-densities $0.08 \leq p \leq 0.22$ (±0.01), each within a 40 nm square field of view and, in total, comprising more than $10^6$ individual g($r$,V) spectra. Our objective is to use this comprehensive data set to explore the evolution with doping of the electronic excitation spectra.

In s-wave superconductors, an increasing quasiparticle inelastic scattering rate reduces their lifetimes and eventually destroys the superconductivity[26]. The signature of this process is manifest in g(V); at zero temperature and with no scattering, two 'coherence' peaks in g(V) occur as singularities on either side of an empty gap and, as scattering rates increase, these peaks decrease in height and increase in width with a rapid increase of the density of excitations near E=0. Such g(V) spectra can be very successfully parameterized by adding an imaginary term Γ to the quasiparticle energy E so that N(E) takes the form[27]

$$N(E,\Gamma) = A \times \text{Re}\left(\frac{E + i\Gamma}{\sqrt{(E + i\Gamma)^2 - \Delta^2}}\right) \quad (1)$$

Here Γ represents a constant scattering rate for quasiparticles. As Γ is increased keeping Δ constant, the coherence peaks diminish, the peak-peak measure of the energy-gap becomes less well defined, and there is a rapid increase of N(0) – all in excellent agreement with the experimentally observed effects in g(r,V).

Our goal is to extend this approach to the cuprate excitation spectra. The N(E) we propose is (at least formally) a natural extension of Eqn. 1

$$N(E,\Gamma_2) = A \times \text{Re}\left(\left\langle \frac{E + i\Gamma_2(E)}{\sqrt{(E + i\Gamma_2(E))^2 - \Delta(k)^2}} \right\rangle_{fs}\right) + B \times E \quad (2)$$

Here $\Delta(\vec{k}) = \Delta_1(Cos(k_x) - Cos(k_y))/2$ is a d-wave energy-gap. The $\Gamma_2(E) = \alpha E$ term represents an effective scattering rate that is linear in energy. In Eqn. 2, A is a normalization factor and B a linear asymmetry term to deal with the ubiquitous background slope of g(V) of Bi-2212. The real part of Eqn. 2 then represents the N(E) function we fit to each measured g(V) (its exact form is determined over the appropriate Fermi surface at each doping[28] - see supplementary materials). Figure 1 shows examples of the N(E) calculated from Eqn. 2 as α increases ($\Delta_1$ remaining constant). We see that the coherence peaks are rapidly suppressed but, because $\Gamma_2(0)=0$, a V-shaped gap is retained for all scattering rates. This is crucial for the successful parameterization of all g(E) since, throughout the Bi-2212 phase diagram, such characteristics are ubiquitous.

We use data sets consisting of atomically resolved and registered g(**r**,V) maps spanning the range of doping $0.08 \leq p \leq 0.22$ (as determined from $T_c$=95K × (1-82.6 (p-0.16)$^2$) ). Their spectra change continuously from quite small gaps ($\Delta_1$~10 meV) with sharp coherence peaks, to large ($\Delta_1$~65 meV) gaps where the vestigial coherence peak can just be resolved[29], to the V-shaped gaps with no apparent coherence peaks that predominate below $p$~10% [Ref.'s 1, 30]. To complicate matters further, at each doping there is a distribution in excitation spectra associated with the distribution of non-stoichiometric oxygen dopant atoms[31], with the probability of these different spectral types varying with doping[4, 29,30,31]. Fitting Eqn. 2 to all these spectra is designed to yield quantitative values for both $\Gamma_2(E)$ and $\Delta_1$ - even when there are no coherence peaks visible and despite both the electronic disorder and the rapid changes in spectral types with doping.

Figure 2 shows the distribution of spectral types[29,30] from within a single field of view, each curve being offset vertically for clarity. The open circles represent the average g(V) spectrum associated with each energy-gap magnitude - the error bars showing the one-$\sigma$ variations of each distribution (see supplementary materials). This averaging process is designed to yield the characteristic excitation spectrum associated with each energy-gap maximum while minimizing complications from the spatial variations in g($r$,V). We emphasize, however, that our fits of N(E) are to each individual local g($r$,V) spectrum (see supplementary materials). The solid lines in Fig. 2 show the average of the fits of Eqn. 2 to the g(V) data – again with all N(E) exhibiting the same $\Delta_1$ averaged together. It is striking how well a very wide variety of g(V) spectral shapes, ranging from those with sharp d-wave coherence peaked spectra to those with V-like spectra having no apparent coherence peaks, can be fitted using Eqn. 2. The fit-quality parameter is a normalized $\chi^2 < 0.01$ for more than 90% of the spectra $0.1 \leq p \leq 0.22$. For the sixth sample with $p$~0.08, the normalized $\chi^2$ remains higher because the strong tunnelling asymmetry[1] prevents good fits. And for $p$>0.22 the spectral shape begins to change in a fashion not yet understood. Nevertheless the vast majority of measured g($r$,V) spectra for 8%<$p$<22% can be fitted very well (a normalized $\chi^2 < 0.01$) using Eqn. 2 . We show in the supplementary materials typical examples of the fit for each value of $\Delta_1$.

In previous studies of nanoscale electronic disorder in Bi-2212 a local energy-gap maximum $\Delta_{pp}$ was defined as half the energy difference between two particle-hole symmetric peaks in g(V) (wherever such pairs of peaks existed). Figure 3, Column 1 shows the spatial and doping dependence of such $\Delta_{pp}$ maps (all FOV are 40nm square and all gap scales are the same with white indicating an inability to measure $\Delta_{pp}$ because coherence peaks could not be identified in high gap regions[29,30]). Figure 3, Column 2 shows the spatial and doping dependence of $\Delta_1$-maps calculated from fits of Eqn. 2 to the identical data sets. We see immediately that the $\Delta_1$-maps resemble closely the $\Delta_{pp}$ –maps. Furthermore, the normalized cross-correlation[31] between all simultaneous pairs of $\Delta_1$-maps and $\Delta_{pp}$-maps shown exceeds 0.9 (where identical images would yield 1). These correspondences between Figure 3, columns 1 and 2 give strong confidence that the Eqn. 2 fitting scheme is working well since the mathematical procedures to make the two kinds of maps are completely different.

New information is immediately available from measurements of the gap maximum $\Delta_1$. A limitation of previous studies was that, when there were weak or no coherence peaks at low doping, it became virtually impossible to determine $\Delta_{pp}$ (such areas were represented in black in Ref.'s 29,30,31 and white in Fig. 3, Column 1). But Fig. 2 shows clearly that with strong effective scattering rates $\Gamma_2(E)$, the coherence peaks should disappear and the density of excited states should appear as a V-shaped spectrum. Therefore $\Delta_1$ can now be extracted in regions where previously it would have been considered unknown. For example, in Fig. 3, Column 2 the black regions now represent measured values $\Delta_1$ rising to above 100 meV in small nanoscale patches at our lowest dopings. The extracted values of $\Delta_1$ (Fig. 3,4) follow the doping dependence of PG energy scale[5,6,7]. Moreover, we find no distinction in terms of the fitted form of N(E) between excitations to the PG energy scale at low dopings, and the familiar excitations of the superconducting state[2,3,4] at higher dopings and lower energies.

Based on accurate mapping of $\Delta_1$ (e.g. Fig. 3, Column 2) we can also examine the doping dependence of electronic disorder for the PG energy scales. In Fig. 4a-f we show these $\Delta_1$-maps, but now each is normalized to the mean value of $\Delta_1$ from that same map and shown using the same colour scale. Remarkably, one cannot distinguish which doping is represented by the images in Fig 4a-f. In Fig. 4g we show the histograms of $\Delta_1/\overline{\Delta}_1$ from these images; it is immediately obvious that the distributions are virtually independent of doping. This indicates that the nanoscale trigger for energy-gap disorder is universal (as it should be for disorder from interstitial substitutions and dopant atoms[31]). Furthermore, since the same fractional distribution about the mean gap-energy is observed for PG energy scales at the low dopings (as $T_c \rightarrow 0$), the PG excitations[32] appear susceptible to the identical nanoscale disorder to those of the superconductor[4,29,30,31].

Next we focus on the most significant discrepancies between fits to Eqn. (2) and the related g(*r*,V) data. These always occur predominantly at the "kinks" which have been reported ubiquitously[1,29,30,31,32,33] in cuprate STM spectra. In general, these kinks are weak perturbations to N(E) near optimal doping, becoming more clear within nanoscale regions increasing in number as *p* is strongly diminished[29,30]. In Figure 5a we show representative $\Delta_1$-sorted spectra. Notice that it is for $\Delta_1$>50meV (with equivalent data for all dopings shown in the supplementary materials) the kinks become more obvious. Each kink is identified by finding the point of inflection as the minimum in the next derivative $d^2I/dV^2$ as shown in Fig. 5b; its energy is labelled $\Delta_0(r)$. We emphasize that these kinks are weak departures from the fits to N(E) (see Supplementary Materials Fig. 4). For the higher energies approaching $\Delta_1$ which are the focus of our study, the kinks neither spoil the excellent fit quality nor the extracted $\Gamma_2(E)$ (see Supplementary Materials Fig. 2). Simultaneous $\Delta_1(r)$ and kink-energy $\Delta_0(r)$ maps can then be derived and are shown in Fig 5c,d. By imaging $\Delta_0(r)$ for all dopings, we find that the excitations are always divided into two categories: E<$\Delta_0$ excitations are homogenous in *r*-space and well defined d-wave quasiparticle eigenstates in *k*-space[34,35], while for E>$\Delta_0$ they are heterogeneous[29,30,31,32,33] and ill-defined in *k*-space. Thus <$\Delta_0(r)$> represents the average energy scale separating spatially homogenous from heterogeneous excitations.

Another new finding involves the capability to estimate local effective scattering rates. Images of the coefficient $\alpha(\vec{r})$, the linear coefficient of the energy dependence of $\Gamma_2$, are shown in Fig. 3 Column 3. These $\alpha(\vec{r})$ range from 0.0 to 0.4 (yellow to black) with the spatially averaged value $<\alpha(\vec{r})>$ growing with falling doping. The scattering rates $\Gamma_2^*$ at $E=\Delta_1$ are most physically significant. These are determined from $\Gamma_2^*(\vec{r}) = \alpha(\vec{r})\Delta_1(\vec{r})$ and are shown in Fig. 3 column 4; they range from yellow (weak scattering) to black (strong scattering) with the maximum effective scattering rates $\Gamma_2^* > 25$ meV for $p \leq 10\%$. One can see the direct correspondence between both the coefficients $\alpha(\vec{r})$ and $\Gamma_2^*(\vec{r})$ with $\Delta_1(\vec{r})$ by comparing Columns 2, 3 and 4. From these, it appears that the relationship between $\Delta_1(\vec{r})$, $\alpha(\vec{r})$ and $\Gamma_2^*(\vec{r})$ is intrinsic and local at the nanoscale.

In Figure 6a we show the value of $\alpha$ associated with each value of $\Delta_1$ throughout six samples with different hole-densities. Overlaid on these data as solid black circles are the $<\alpha(\vec{r})>$ versus the spatially average $<\Delta_1>$ for each sample; they are in good agreement with the relationship between local pairs of $\Delta_1$ and $\alpha$ values throughout. These data demonstrate that the relationship between $\Delta_1(\mathbf{r}):\alpha(\mathbf{r})$ pairs is local at the nanoscale and apparently intrinsic - since it is the same in all samples at all dopings. Again, we conclude that whatever electronic process perturbs the energy-gap distribution[29,30,31,32,33,36] perturbs the effective scattering rate $\Gamma_2(E)$ locally in a related fashion.

Significant new insights emerge from these fits when summarized in the form of a phase diagram. In Fig. 6b we show $<\Delta_1>$ as blue circles; it rises linearly with decreasing $p$ along the well-known[5,6,7] trajectory for excitations to the PG energy scale. The black circles represent the spatially averaged $E=\Delta_1$ scattering rates $<\Gamma_2^*>$; these are very low when $p>16\%$ but undergoes a strong transition to a steeply rising trajectory for $p<16\%$. This dramatic increase of the effective scattering rates for states away from the nodes, culminates in another transition somewhere below $p\sim10\%$ with the appearance of extreme tunnelling asymmetry[1,30] (rendering efforts to fit Eqn. 2 impossible). Finally, the red circles represent the spatial average of the second energy scale $\langle\Delta_0\rangle$ where both the ubiquitous 'kink' occurs in the g($r$,V) spectrum, and above which spatial homogeneity in quasiparticle excitations is lost. Clearly $\langle\Delta_0\rangle$ diverges from $\langle\Delta_1\rangle$, falling slowly as $p\rightarrow 0$.

*Discussion/ Conclusions*

In this paper we introduce a new technique for analyzing the tunnelling-derived cuprate electronic excitation spectrum N(E) as T→0. The results provide a significantly more quantitative and comprehensive picture of the T→0 excitations than was previously available and for a wide range of hole-densities. And, since this fitting technique is demonstrably successful under a very wide variety of circumstances, we can also anticipate its extension to new arenas such as at high temperatures[32] or when additional phase fluctuation effects occur near vortex cores[4]. It is important, however, to be aware of the limitations of any interpretation of $\Gamma_2$(E) simply as a one-particle scattering rate. Equation (2) might be taken as an expression for a classic d-wave superconductor with single-particle scattering rate $\Gamma_2$ within BCS theory. Such an interpretation, which may possibly be appropriate in the overdoped materials, would assume weakly interacting quasiparticles. But as the Mott insulator is approached at strong underdoping, this intrinsic effective scattering rate may become so intense that such single-particle $k$-space excitations are no longer well-defined even in the superconducting state (especially near the Brillouin zone face)[8,9,21,23,25]. Only a few authors have investigated theoretically the lifetime of such quasiparticles in the superconducting state and away from the nodes. Spin fluctuation theories of d-wave superconductivity suggest relatively weak dependence of the scattering rate on the direction of the quasiparticle momentum[37,38]. In the underdoped cuprates, pair breaking scattering from vortex-antivortex pairs has been proposed as the origin of large ARPES spectral widths near the antinode[39,40]. Another caution about the effective scattering rate $\Gamma_2$ discussed here is that it is related to the *local* Green's function G($r,r$), whose spectral characteristics will be broadened by scattering processes involving the entire Fermi surface. It is then far from clear that a general fit of the form of Eqn. 2 with a local self-energy should obtain; in an inhomogeneous system, the self-energy is a bilocal quantity $\Sigma(\mathbf{r},\mathbf{r}')$. Our findings that the vast majority of spectra can be fit, at least for E>$\Delta_0$, by an identical form as Eqn. 2 and that $\Gamma_2(\mathbf{r})$ is spatially correlated with order parameter $\Delta_1(\mathbf{r})$, imply that $\Gamma_2(\mathbf{r})$ does represent the effective 'local' self-energy of a quasiparticle sampling a region of size less than or equal to the gap "patch" size, i.e. that the system is self-averaging on this scale. A final caveat is that $\Gamma_2(E) = \alpha E$ represents the first approximation to the true energy dependence of scattering rates consistent with the spectra; it captures very well the low scattering of near nodal quasiparticles and the intense scattering $\Gamma_2^*$ at E=$\Delta_1$. Eventually, however, a more complex form for $\Gamma_2(E)$ consistent with everything reported herein but capturing finer details of changes in scattering rate throughout $\mathbf{k}$-space may be required.

Nevertheless, a number of important conclusions result from these data and fitting procedures. Local quasiparticle lifetimes $\tau \sim 1/\langle\Gamma_2(E)\rangle$ can now be determined from STM data. If we focus on $\langle\Gamma_2\rangle$ as a function of $p$, we find a very distinct change near optimal doping characterised by appearance and extremely rapid growth of scattering rates towards the underdoped regime. This latter effect signifies such intense scattering near the antinodes at lowest dopings, that it must be closely related to the disappearance[8,9] of well defined $\mathbf{k}$-states there. Moreover, $\Delta_1(\mathbf{r})$ and the coefficient of energy dependence in the effective scattering rate $\alpha(\mathbf{r})$ appear to be linked intrinsically and locally - retaining the same relationship throughout all samples. The rapid increase of $\Gamma_2^*$ scattering rates as the Mott insulator state is approached is likely due to electron-electron interactions but the exact microscopic processes cannot be identified from this study. Significantly, we find no apparent distinction in terms of the form of N(E) in Eqn. 2 between fits to optimally doped g(V) spectra which definitely represent d-wave superconductivity, and the g(V) spectra of strongly underdoped samples down to $p$~10% as superconducting $T_c$ diminishes towards 0. This means that a combination of a single particle-hole symmetric energy gap $\Delta(\vec{k}) = \Delta_1(Cos(k_x) - Cos(k_y))/2$ plus an effective scattering rate $\Gamma_2(E) = \alpha E$ provides a good description of excitation spectra - without recourse to a distinct electronic ordered state. We emphasis that these conclusions might not hold at $p<10\%$ because spectra are no longer well fit by Eqn. 2 due to strong tunnelling asymmetry[1]. Furthermore, our results for $p>10\%$, do not imply that there is only one energy scale present: consistent with both the wide variety of long-standing results[6,7, 9,25,29,30,33] and the more recent spectroscopic observations[41,42,43], we find that two energy scales always exist on the underdoped side of the phase diagram. The higher scale $<\Delta_1>$ evolves along the PG line. Here we find that the lower scale $<\Delta_0>$, representing segregation in energy between homogenous and heterogeneous electronic structure, diverges from $<\Delta_1>$ when the $\Gamma_2$ scattering rates begin to increase rapidly.

An intriguing scenario stimulated by these observations would be that superconducting cuprates exhibit an overarching d-wave energy scale $\Delta_1$ but the related quasiparticles experience rapidly increasing anisotropic scattering rates as $p\rightarrow 0$. This scenario has recently become the focus of intense theoretical study[44] yielding a number of far reaching conclusions including (i) realistic calculations of impurity- and spin-fluctuation scattering contributions to local density of states showing that typical quasiparticle scattering rates are indeed quasi-linear in energy and proportional to $\Delta_1$, (ii) demonstration of how the mean free path falls drastically with increasing quasiparticle energy so that, below a critical bias, all quasiparticles explore so many heterogeneous gap patches that their spectra appear homogeneous (iii) evidence that that the quasiparticle interference peaks[34,35] could be weakened primarily by inelastic scattering represented by $\Gamma_2(E)$ and, (iv) reconciliation of photoemission with STM tunnelling and neutron-scattering lifetimes, by inclusion of gap inhomogeneity-induced broadening of ARPES spectral function. Moreover, quasiparticles subject to scattering rates above some critical value of $\Gamma_2(E)$ might not retain sufficient coherence to contribute to the superfluidity in the ground state[45] thus leading to the ultimate breakdown of superconductivity as $\Gamma_2^*$ diverges at low doping.

To test these new hypotheses will require (i) determination of whether the superconducting quasiparticles are actually governed by a pairing gap on the scale of $\Delta_1$ as $p \to 0$, and (ii) microscopic identification of the $\Gamma_2$ scattering process and its relationship to $\Delta_0$ where energy-segregation of homogenous from heterogeneous electronic structure begins

*Acknowledgements*

We acknowledge and thank D. Dessau, P. Johnson, E.W. Hudson, D. H. Lee, P.A. Lee, A. P. Mackenzie, A. Millis, M. Norman, N. P. Ong, M. Randeria, D. J. Scalapino, K. Shen, T. Timusk, Y. Uemura and T. Valla for helpful conversations and communications. This work is supported by NSF through the Cornell Center for Material Research, by the Cornell Theory Center, by Brookhaven National Laboratory under Contract No. DE-AC02-98CH1886 with the U.S. Department of Energy, by U.S. Department of Energy Awards DE-FG02-06ER46306 and DE-FG02-05ER46236, by the U.S. Office of Naval Research, and by Grant-in-Aid for Scientific Research from the Ministry of Science and Education (Japan) and the 21st-Century COE Program for JSPS. Fellowship support is acknowledged by K. F from I$^2$CAM.

*Figures*

Figure 1: Representative N(E) from Eqn. 2 demonstrating the effect of increasing $\alpha$ for $\Delta_1$=20meV. The black line represents $\alpha$=0.00, the red line for $\alpha$=0.05, the green line for $\alpha$=0.10 and the blue line for $\alpha$=0.40.

Figure 2: Open circles represent the average value of g(V) from all spectra (in one sample with *p*=10%) that exhibit a given gap magnitude $\Delta_1$. The error bars give one standard deviation of the distribution in g(V) at each V. The corresponding average of the fits of all spectra by Eqn. 2 are shown as solid lines. The table shows the fitted values of $\Delta_1$ and $\Gamma_2^*=\Gamma_2$ (E=$\Delta_1$). The $\Delta_1$ ranges from 38mV to 93mV while $\Gamma_2^*$ spans from below 1meV to above 25meV. Each spectrum is offset for clarity. Notice the particle-hole symmetry throughout.

Figure 3: Column 1 (a-e) shows the $\Delta_{pp}$ maps as a function of doping - each for a 40nm$^2$ field of view, the white areas are places where $\Delta_{pp}$ cannot be defined. The dopings are calculated from the T$_c$'s of the samples using the formula T$_c$=95K × (1-82.6 (p-0.16)$^2$) and are (a) 0.22±0.01, (b) 0.19±0.01, (c) 0.17±0.01, (d) 0.14±0.01 and (e) 0.10±0.01. Tunnelling asymmetry renders fitting the sixth data set at *p*~8% very difficult. The $\Delta_1$ maps calculated from the fits to Eqn. 2 using the identical original g(*r*,V) maps as in Column 1, are presented in Column 2 (f-j). Note that where $\Delta_{pp}$ and $\Delta_1$ can both be evaluated they create virtually identical patterns. Column 3 (k-o) shows the $\alpha(\vec{r})$ calculated concurrently with each $\Delta_1$ from the fits to Eqn. 2. Column 4 (p-t) shows the corresponding maximum effective scattering-rate maps $\Gamma_2^*$, calculated from Column 2 and 3. Note that $\Delta_1$, $\alpha$, and $\Gamma_2^*$ create very similar patterns. T$_c$ for each sample is shown as inset to the left hand panels.

Figure 4: (a-f) Normalized $\Delta_1$ maps for six hole-densities 0.8<p<0.22 for 40nm$^2$ g(*r*,V) data sets,. The dopings are (a) 0.08, (b) 0.10, (c) 0.14, (d) 0.17, (e) 0.19 and (f) 0.22. The maps were normalized to the average value of $\Delta_1$ value in each g(*r*,V) maps. For *p*=0.08

we can only estimate the value of $\Delta_1$ from fits to the positive bias part of the spectrum where the steep tunnelling asymmetry is less prominent.

(g) Histograms of the data in Fig 4a-f. Obviously, these distributions are statistically highly similar.

Fig 5:

(a) A set of $\Delta_1$-sorted spectra shown with an expanded vertical scale designed to emphasize the representative the kinks occurring ubiquitously in $\Delta_1$>50meV spectra. The red arrow labelled $\Delta_0$ points to average energy at which such kinks are detected in dI/dV.

(b) Energy of each kink $\Delta_0(r)$ is identified by finding the point of inflection as the minimum in the next derivative $d^2I/dV^2$. The black line is the spatially averaged value of $d^2I/dV^2$, the red line is the spatially averaged derivative of the fits to Eqn. 2 and the red arrow labelled $\Delta_0$ indicates the kink energy.

(c) Gap-energy $\Delta_1(r)$ map

(d) Kink-energy map $\Delta_0(r)$ simultaneous with (c). Clearly the kinks are associated with the higher energy gap spectra, and observation found true at all dopings.

Figure 6:

(a) The local relationship between $\alpha(r)$ and $\Delta_1(r)$ using all the N(E) fits for the average hole-densities <p> shown. The spatial average value of <$\Delta_1$> and <$\alpha$> for each of the five different samples is plotted as large coloured circles. The global average relationship between <$\Delta_1$> and <$\alpha$> appears to be indistinguishable from the local relationship between $\alpha(r)$ and $\Delta_1(r)$.

(b) The doping dependence of fitted <$\Delta_1$> (blue circles), <$\Delta_0$> (red circles) and <$\Gamma_2^*$> (black squares) each set interconnected by dashed guides to the eye. The higher scale <$\Delta_1$> evolves along the PG line[5,6,7] while the lower scale <$\Delta_0$> represents segregation in energy between homogenous and heterogeneous electronic structure. The separation of <$\Delta_1$> from <$\Delta_0$> scales begins to occur at the point where <$\Gamma_2^*$> starts to rise rapidly. $T_c$ and p for each sample is shown as inset.

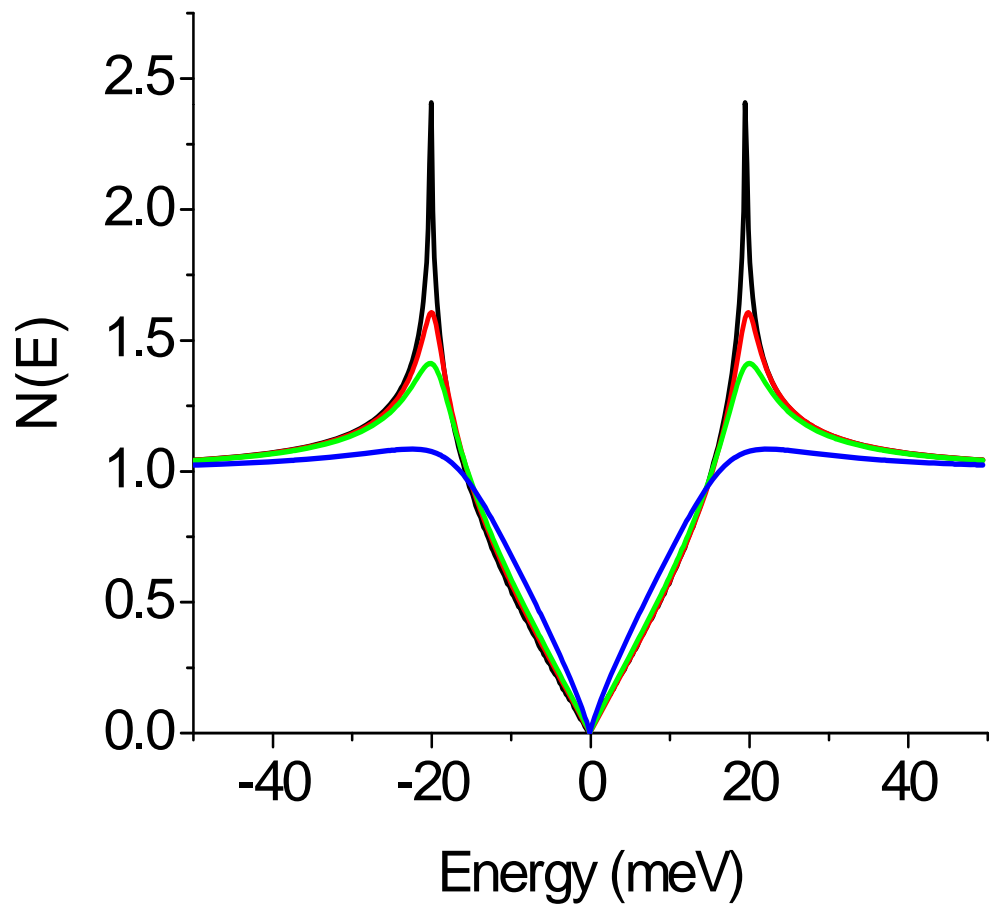

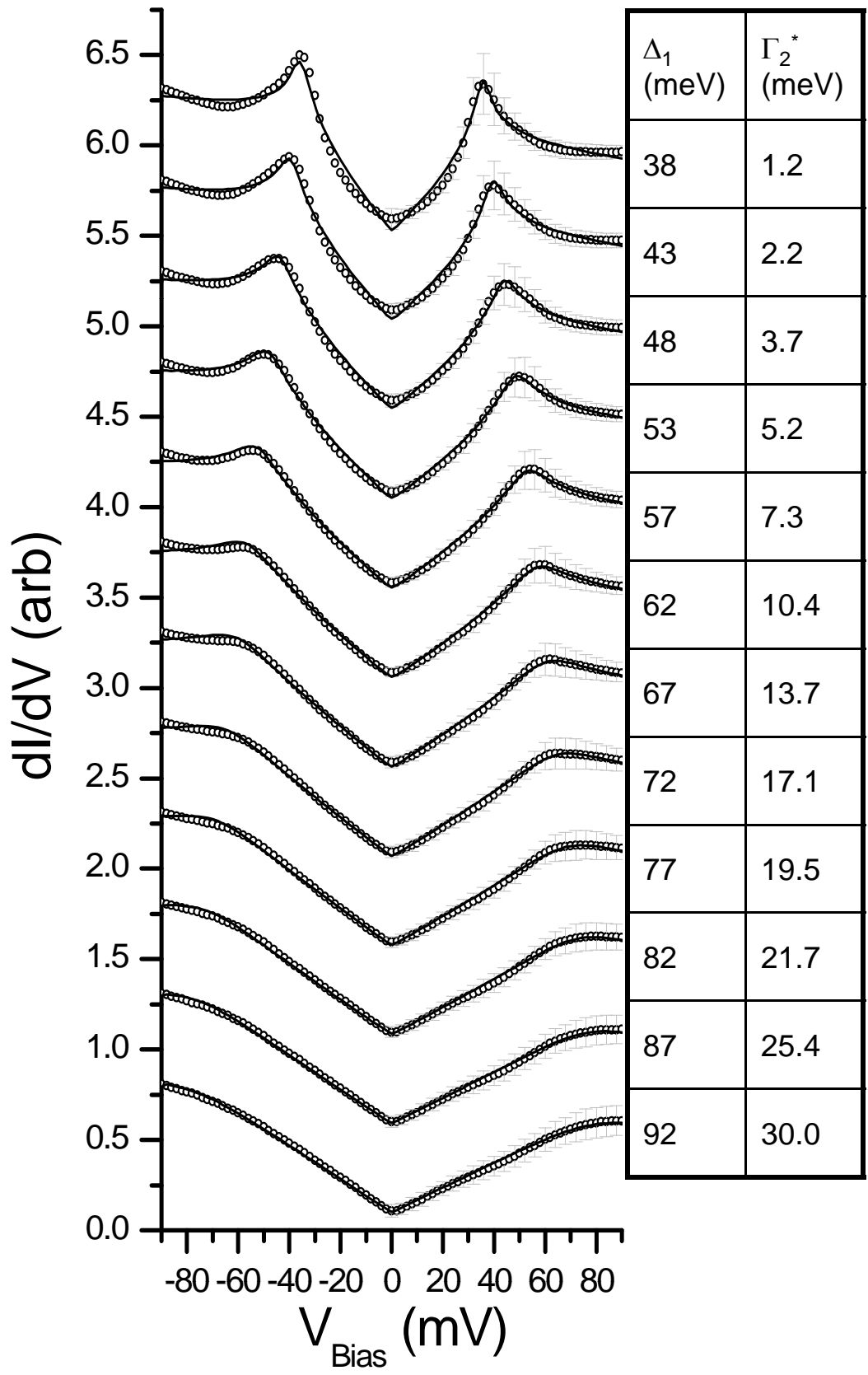

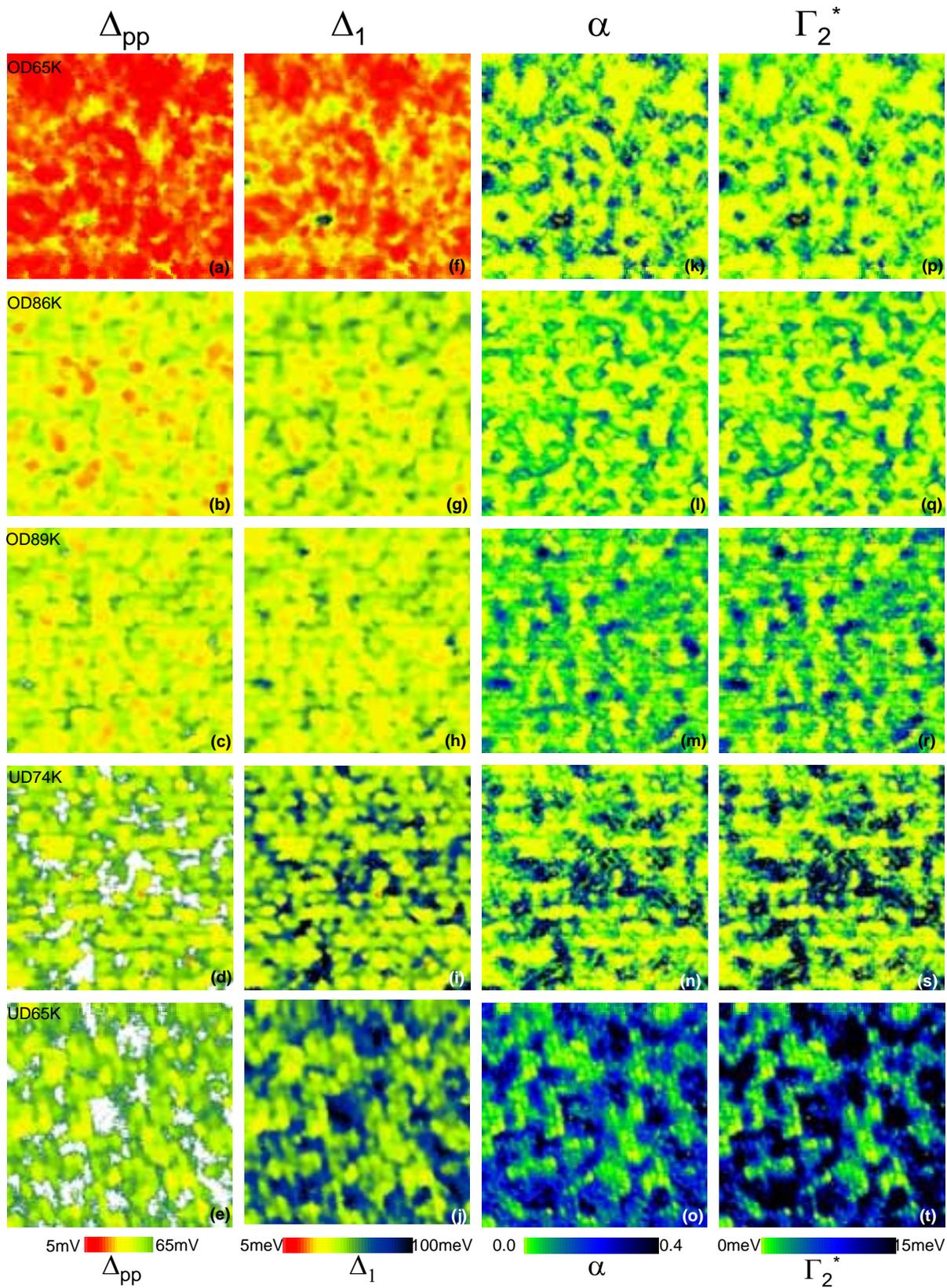

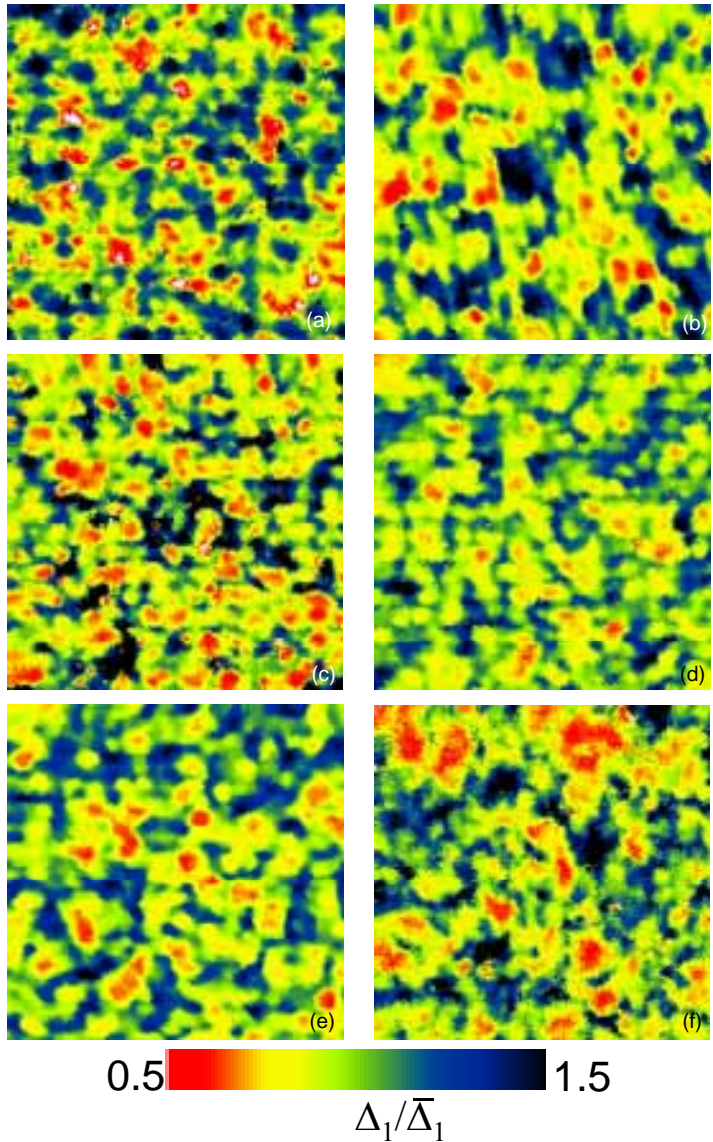
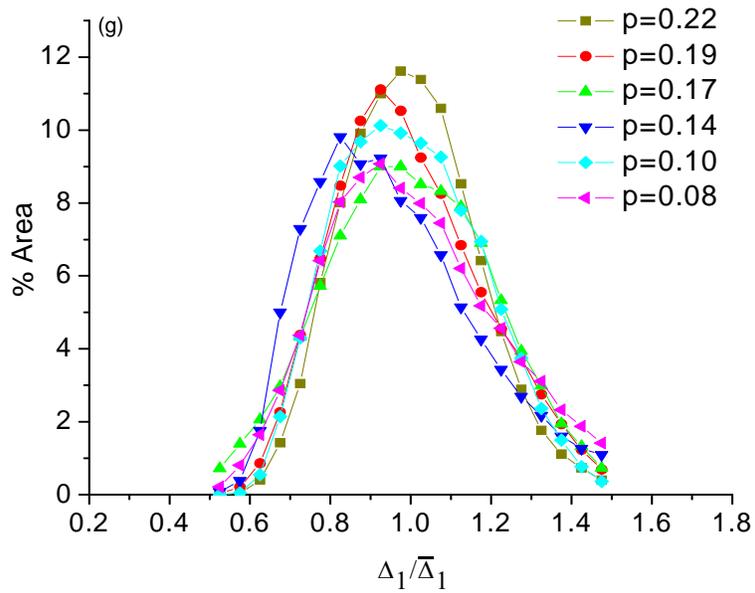

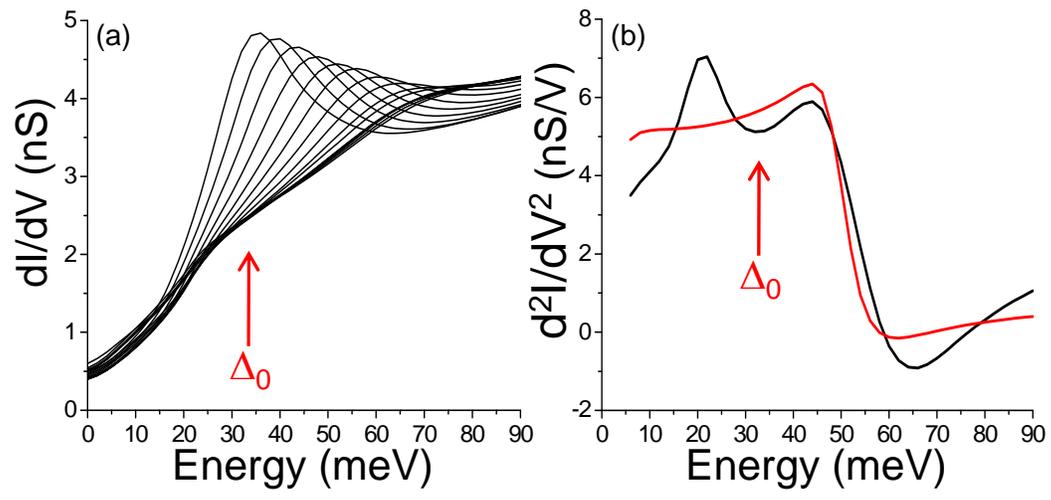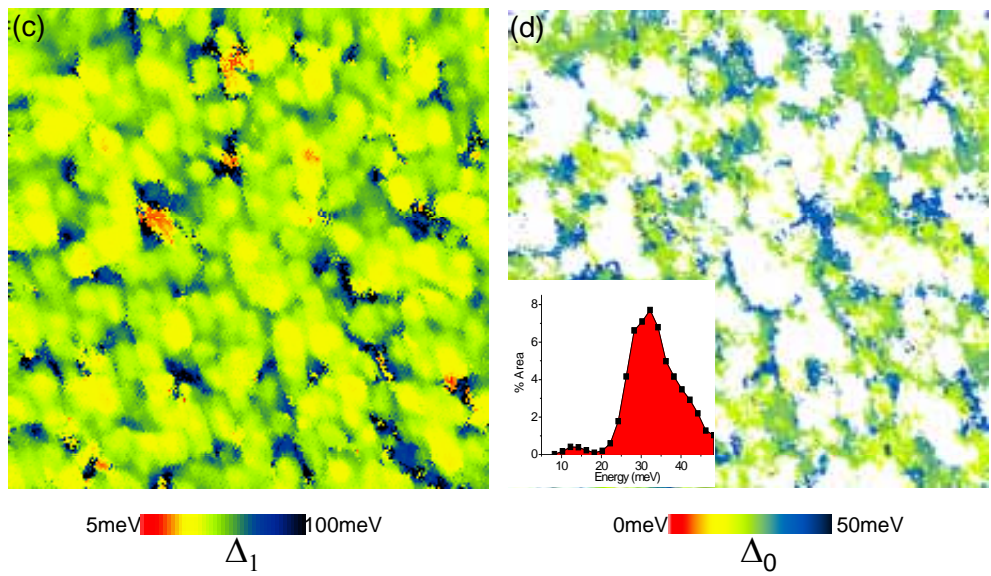

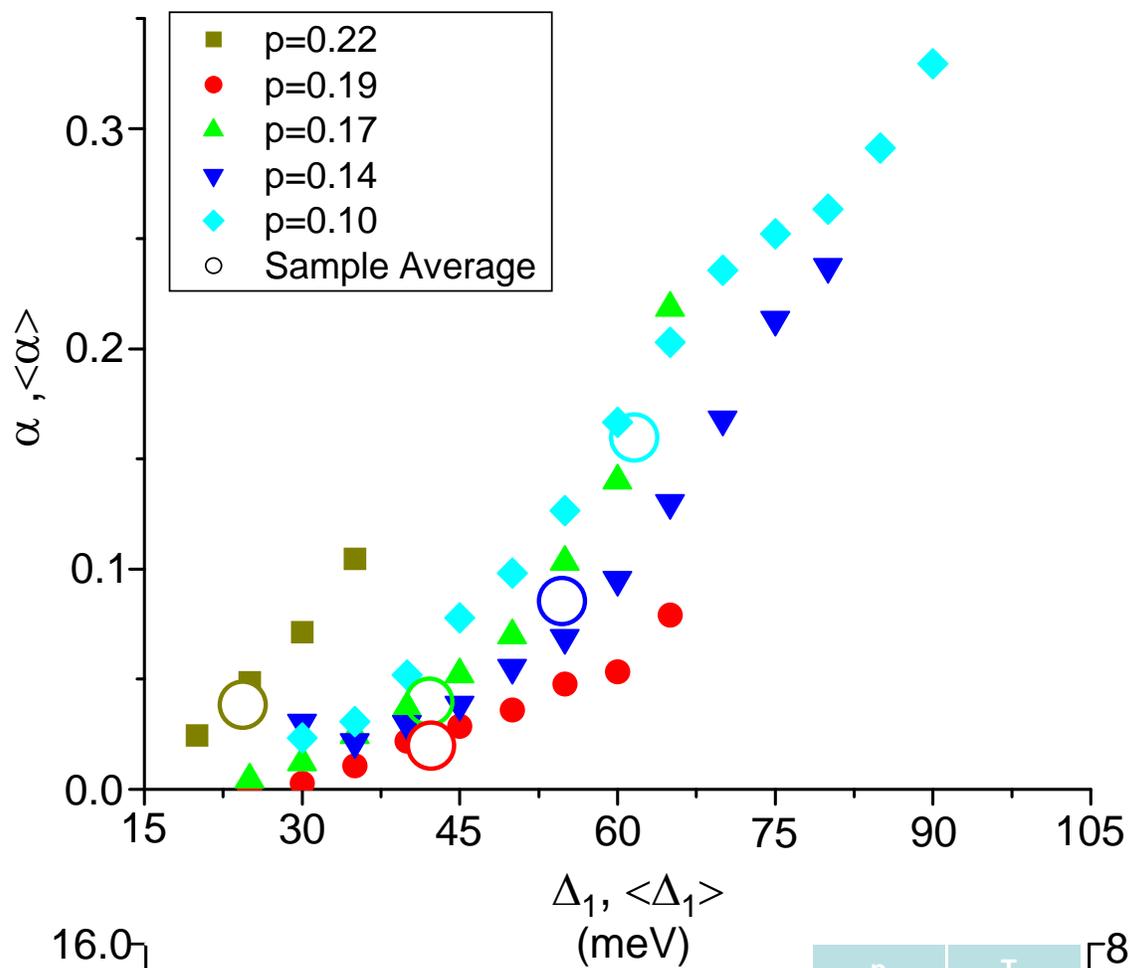
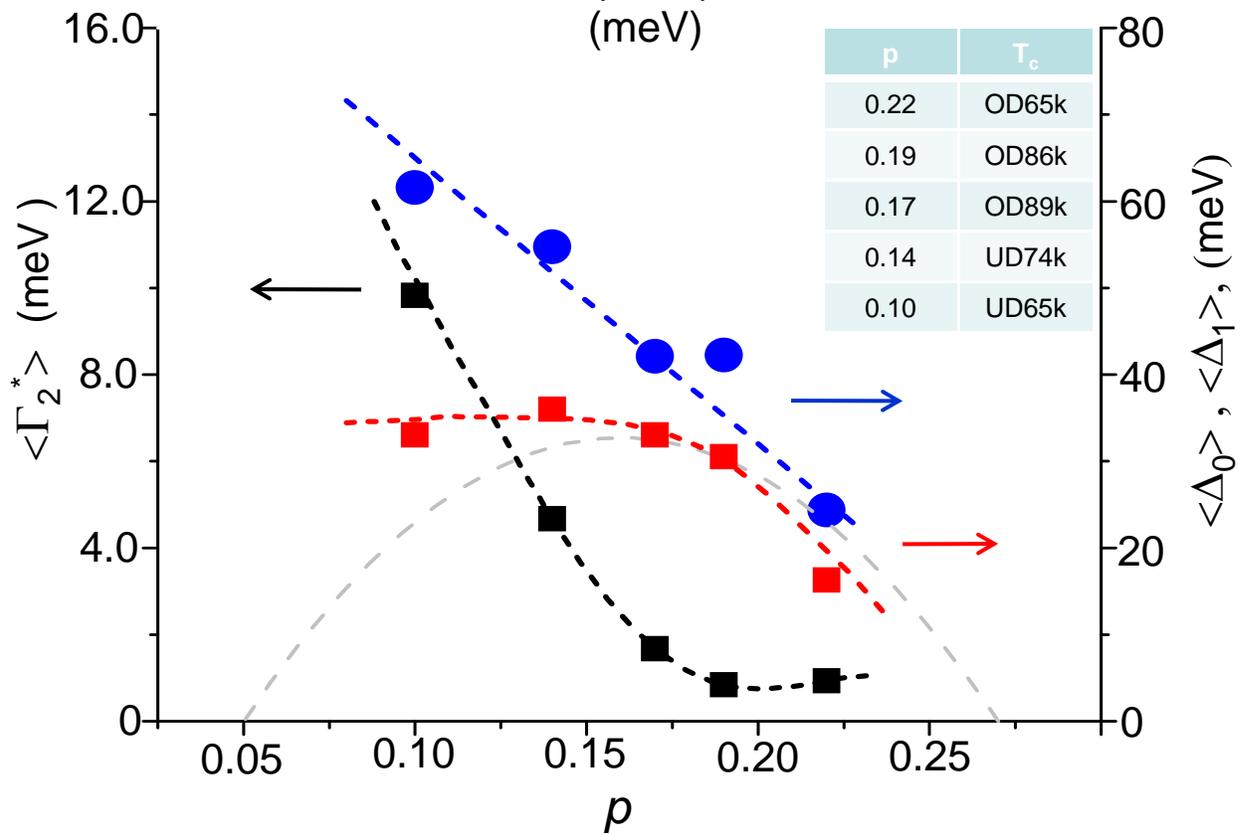

# Supplementary Material

*Supplementary Materials*

While Eqn. 2 appears to be quite simple, it is numerically intensive. We use the Levenberg-Marquardt algorithm implemented in C [1]. The algorithm computes the Jacobian of the function N(E) and looks at the difference between the g(V) and the fit as a function of the small changes in the parameters. With each step, the Levenberg-Marquardt algorithm computes this change and diminishes the difference with iteration.

For a given doping, the Fermi surface is designated using the tight binding model parameterized by Norman *et al*. An array of points equally spaced in distance along the Fermi surface is then calculated for E=0. Equation 2 is evaluated for each of these points, adding the $k_x, k_y$ coordinate into the d-wave gap equation. These results are then numerically integrated along the Fermi surface to yield N(E).

Because of the Fermi surface evaluation, each five parameter fit requires approximately a minute. For roughly $10^6$ curves, approximately two years of processing would then be required. Instead, we use hundreds of processors in parallel – a task made simpler by the fact that each fit is independent from all the rest. A typical field of view (256X256 curves) is broken into 8X8 pieces and distruted to 100-200 processes and allowed to run for 1-3 days. The fitting processes to carry out 5-10 times, with different starting parameters in order to minimize the chances of finding a false local minimum.

The quality of fits of N(E) to each spectrum throughout the data set can be demonstrated in a variety of different ways. We use a normalized $\chi^2$ as a measure of the quality of our fit:

$$\chi^2 = \frac{1}{n-p} \sum_{i=0}^{n} \frac{(x_i - f_i)^2}{x_i}$$

This is the standard $\chi^2$ normalized by the number of points minus the number of fitting parameters to allow us to compare data sets that have a varying numbers of points per curve. Here *n* is the number of points per curve, *p* is the number of fitting parameters, *x* is the measured data value and *f* is the fit value.

---

[1] Lourakis, M.I.A.  http://www.ics.forth.gr/~lourakis/levmar/ (2004)

In Supplementary Figure 1 we demonstrate the fits to N(E) to individual g(V) spectra along with the fit parameters. In Supplementary Figure 2 we show that the histogram of fit quality parameter for a complete data set of >64,000 spectra remains low for all $\Delta_1$. For representative individual spectra, we demonstrate the fit quality represented by these low values of $\chi^2$ and also indicate by a red line the location of the 'kink' as identified by a local minimum in $d^2I/dV^2$. We see directly that each 'kink' is merely a small departure from an overall very successful fit; this is universally true when $\chi^2 < 0.01$. Supplementary Figure 3 we show the $\Delta_1$-averaged g(V) spectra and their $\Delta_1$ averaged N(E) fits for spectra from samples at five different dopings demonstrating how the fit quality is preserved across much of the phase diagram. Supplementary Figure 4a shows gap-averaged spectra and their corresponding fits (Fig. 4b). The derivate of these spectra shows a small dip which is associated with the kink, but the derivative of the fits (Fig. 4d) does not. Thus, overall, the effect of kinks is very weak and only detectable globally because of the excellent quality of our fits to Eqn 2.

Supplementary Figure 1: **Individual spectra and fits taken at ten different points**. These data span a 50meV range in $\Delta_1$. The open circles are the raw data, and the solid lines are the fits to the data. The table on the right shows the fit parameters for each of the curves. This is for *p*=0.10.

Supplementary Figure 2: **Normalized $\chi^2$ histogram with example curves.** (a) A 2D histogram showing the grouping of the normalized $\chi^2$ as a function of $\Delta_1$. The scale stretches from colored (red being the highest number of points) to white (zero points). From this histogram five example curves of quality of fit represented by these $\chi^2 <0.01$ are shown. The red line marked by $\Delta_0$ shows the kink location for each of the example curves. The normalized $\chi^2$ remains extremely low despite the presence of the kink.

Supplementary Figure 3: **$\Delta_1$ averaged spectra and fits taken from 4 different dopings and spanning the range of $\Delta_1$ values.** The error bars are the 1-$\sigma$ distribution for the $\Delta_1$

averaging of the spectra. Spectra, counting down from the top, are 1-3 from *p*=0.22, spectra 4-5 from 0.19, spectra 6-8 from 0.17, and spectra 9-17 from *p*=0.10. The colors correspond roughly to the $\Delta_1$ map color scale in Figure 3.

Supplementary Figure 4: **Expanded view of gap-averaged spectra for $\Delta$>50meV**

(a) Gap-averaged spectra from the complete data set whose individual fit quality is demonstrated in Suppl. Fig. 2, (b) the corresponding gap-averaged fits , (c) shows the derivate of the spectra with the approximant kink energy $\Delta_0$ shown by the red line, (d) presents the derivative of these fits. The departure due to kinks of measured spectra from the fits appears as a small dip in $d^2I/dV^2$ in (c) with no equivalent dip in the fit derivative (d).

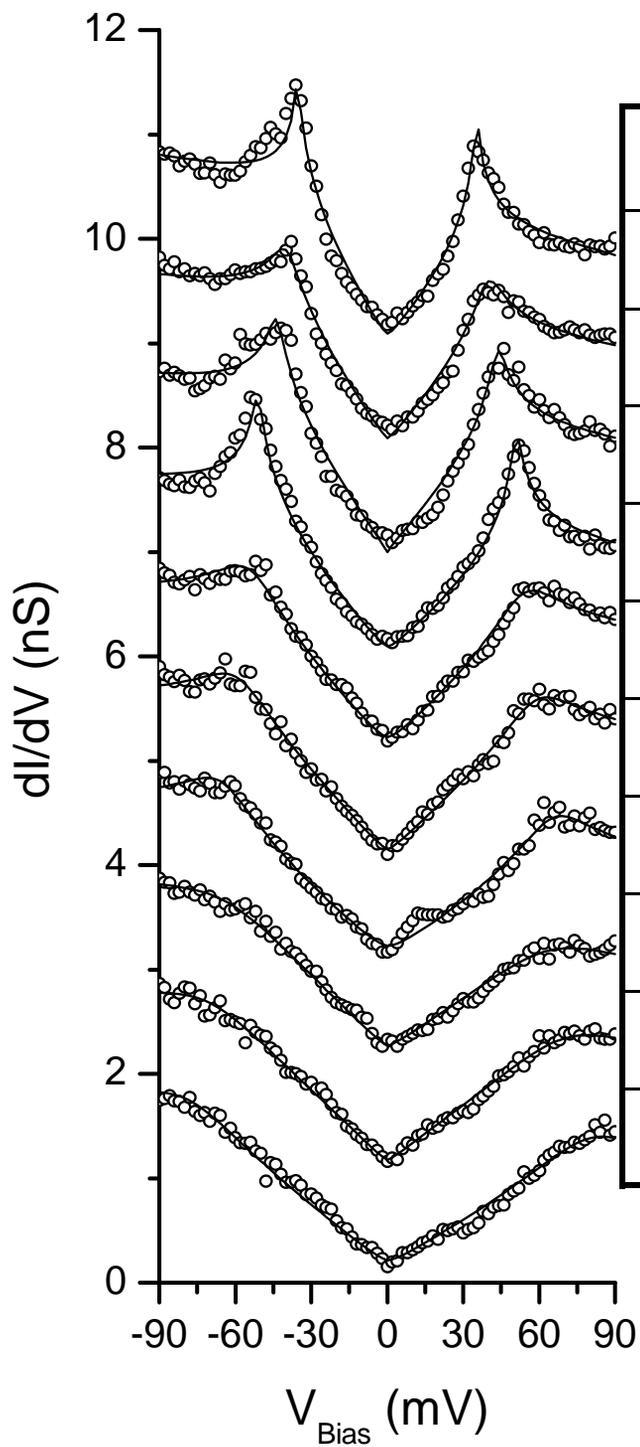

| $\Delta_1$ (meV) | $\Gamma_2^*$ (meV) | $\overline{\chi}^2$ |
|---|---|---|
| 40.0 | 0.11 | 0.010 |
| 42.2 | 3.67 | 0.006 |
| 47.2 | 2.62 | 0.011 |
| 55.9 | 0.30 | 0.008 |
| 60.2 | 7.44 | 0.007 |
| 65.8 | 10.48 | 0.005 |
| 72.3 | 7.24 | 0.006 |
| 76.9 | 24.21 | 0.005 |
| 82.2 | 23.79 | 0.003 |
| 90.8 | 20.78 | 0.006 |

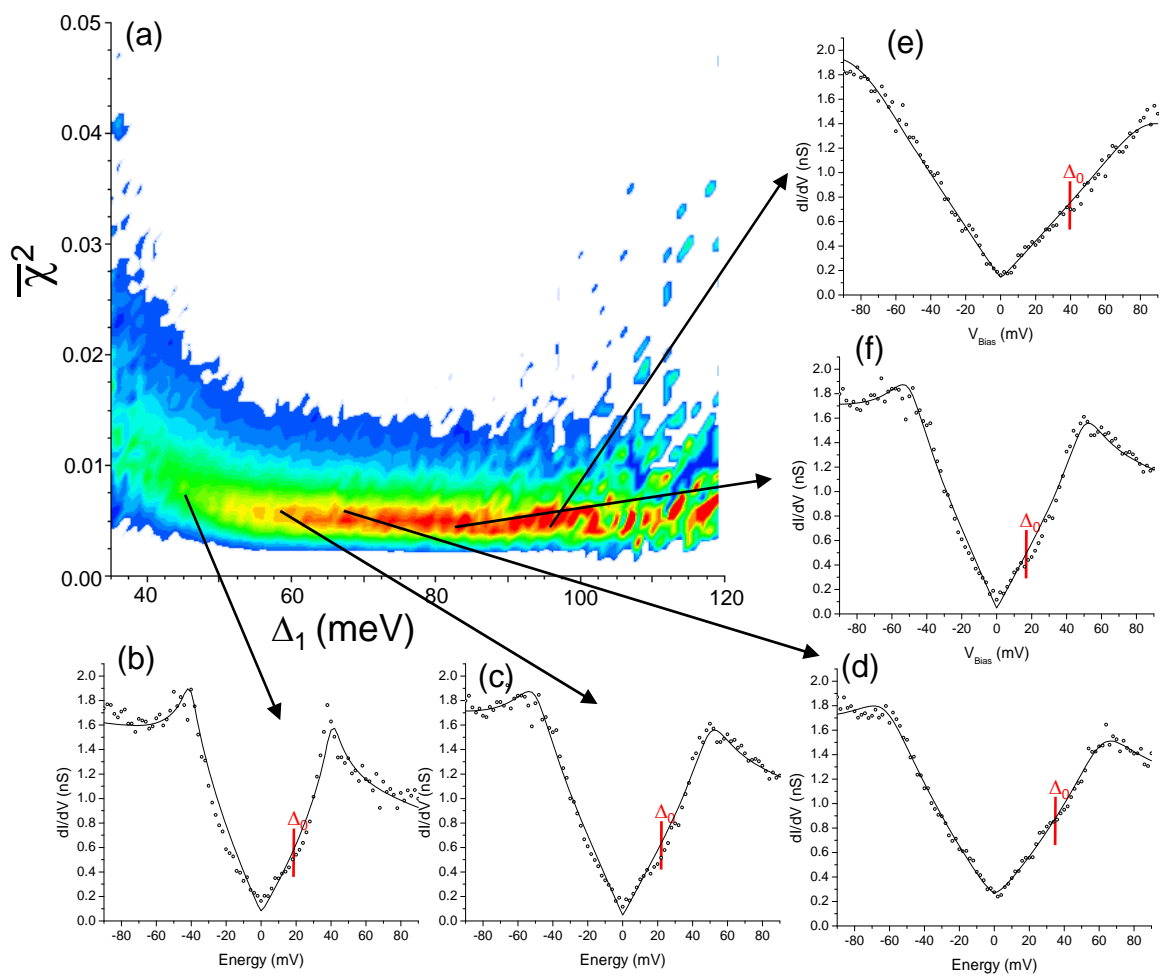

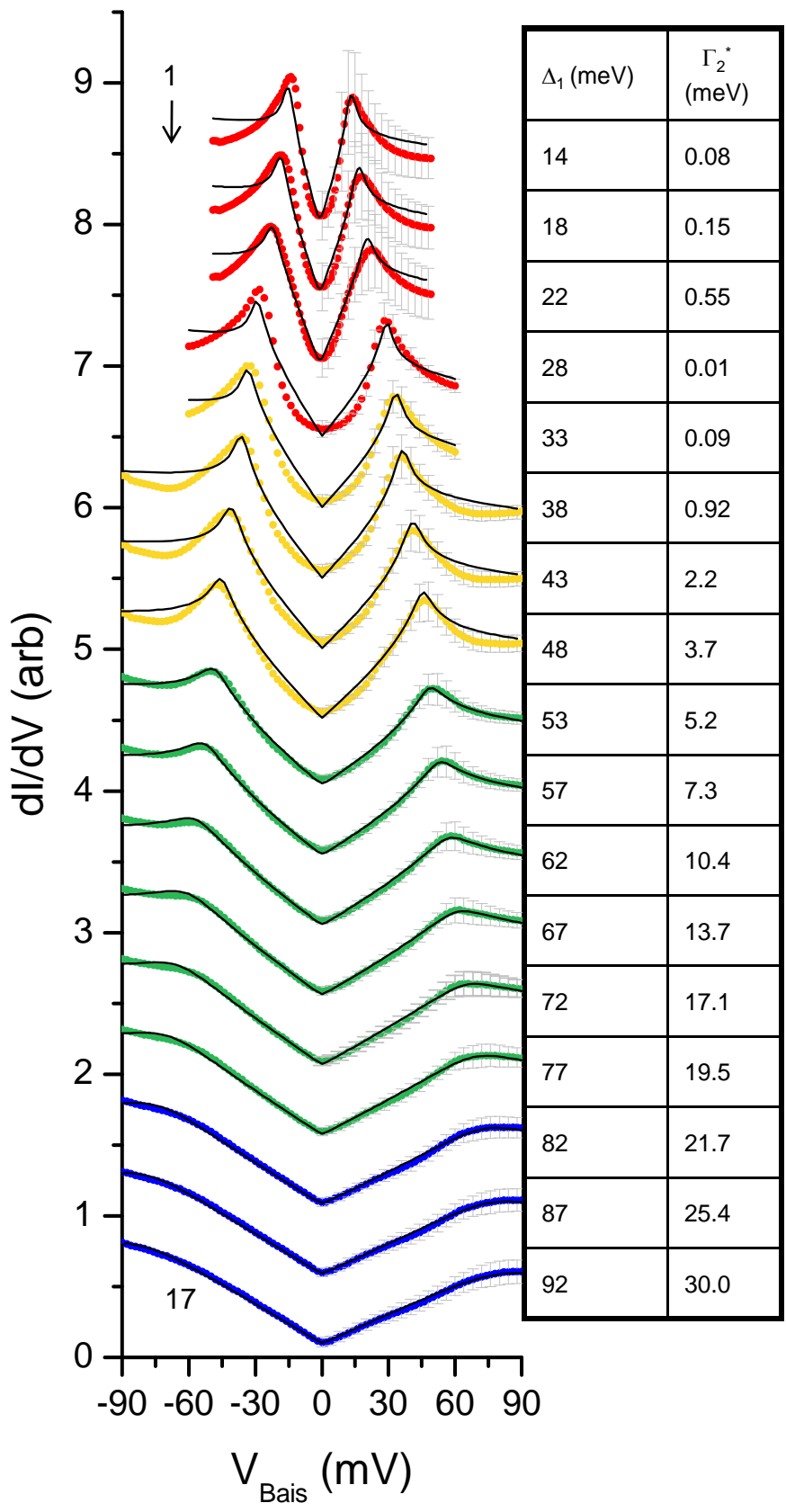

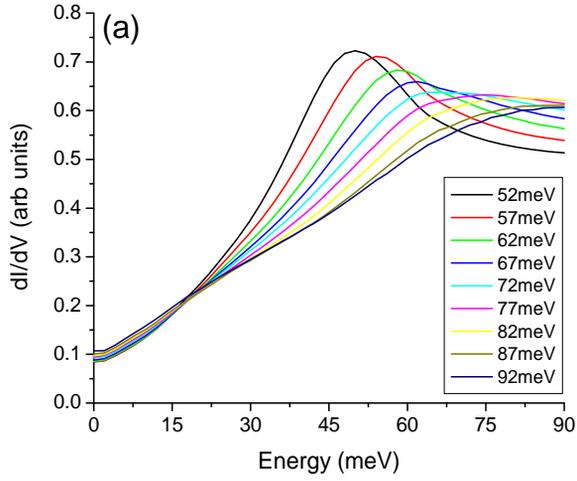
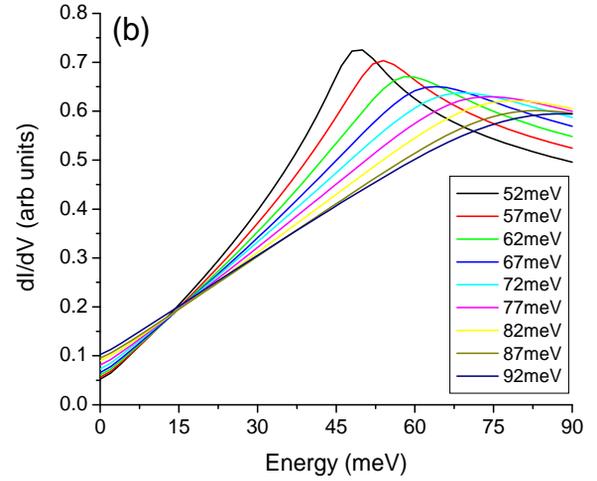
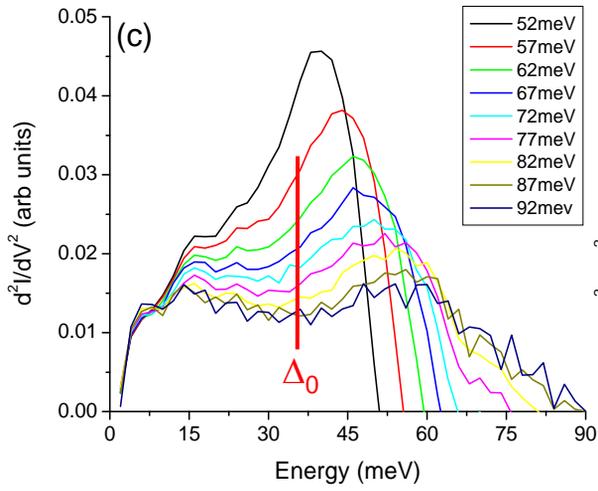
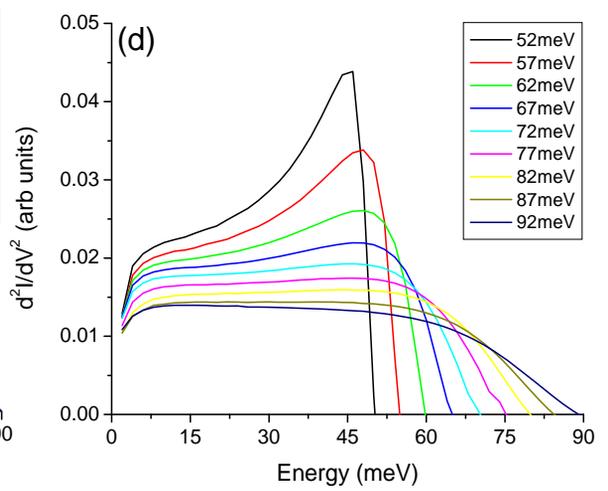